\documentclass{article}
\usepackage{spconf,amsmath,graphicx}

\def\B#1{{\mathbf #1}}

\title{ICA-BASED SPARSE FEATURES RECOVERY FROM FMRI DATASETS
}
\name{Ga\"el Varoquaux$^{1,2}$, Merlin Keller$^{1,2}$, 
Jean-Baptiste Poline$^{2}$, 
Philippe Ciuciu$^{2}$, Bertrand Thirion$^{1,2}$
\thanks{Funding from INRIA-INSERM collaboration.
}}
\address{$^1$ Parietal project team, INRIA, Saclay-\^{I}le de France,
Saclay, France, \\
$^2$ CEA, DSV, I$^2$BM, Neurospin, Saclay, France
}
\begin{document}
%
\maketitle
\begin{abstract}
Spatial Independent Components Analysis (ICA) is increasingly used in the
context of functional Magnetic Resonance Imaging (fMRI) to study
cognition and brain pathologies.
Salient features present in some of the extracted Independent Components
(ICs) can be interpreted as brain networks, but the
segmentation of the corresponding regions from ICs is still ill-controlled.
Here we propose a new ICA-based procedure for extraction of sparse
features from fMRI datasets.
Specifically, we introduce a new thresholding procedure that controls the
deviation from isotropy in the ICA mixing model.
Unlike current heuristics, our procedure guarantees an exact, possibly
conservative, level of specificity in feature detection.
We evaluate the sensitivity and specificity of the method on synthetic and
fMRI data and show that it outperforms state-of-the-art approaches.
\end{abstract}
\section{Introduction} \label{sec:intro}

In neuro-imaging, ICA is the most popular method to explore the
spatial correlation structure of fMRI signals. Some extracted ICs match
well-known brain networks \cite{mckeown1998, beckmann2004} and have been
shown to correspond to units targeted by neuro-degenerate diseases
\cite{seeley2009}. These sources form spatial maps that represent sparse
networks of brain activity: only a small percentage of the voxels
observed are active in a given network. Daubechies {\sl et al.} \cite{daubechies2009}
have argued that this sparsity is key to the success of ICA in the
context of fMRI. When applied to data generated from sparse sources, ICA
amounts to sparse coding \cite{hyvarinen1998}. It has enjoyed more
success in the neuro-imaging community, probably because it groups
together correlated features into components interpreted as brain
networks. Current state-of-the-art ICA models for fMRI (MELODIC
\cite{beckmann2004}) apply univariate mixture models to ICs to separate
signal from noise and recover the sparse structure.

In this paper, we present a multivariate model of sparse brain activity and
an associated procedure for recovering the sparse features with a
statistical control of false detections in the presence of noise. We will
focus on single-subject analysis but the method could easily be extended to
group analysis with the addition of a group model.



\section{Signal modeling and estimation}
\label{sec:model_of_the_signal}

ICA is an unsupervised learning algorithm. As such, it does not provide a
framework for statistical-significance testing, but can be used to analyze
fMRI data without external correlates, such as in resting state. We
introduce a model of the fMRI signal based on the assumption of very
sparse sources.

{\bf Generative model.}
In the observations from the scanner $\B{Y}$\footnote{$\B{Y}$ corresponds
to the data from the scanner after slice-timing interpolation and motion 
correction. In addition, when doing group analysis, a normalization
procedure is often applied, followed by Gaussian spatial smoothing.}, the
underlying BOLD dynamics is confounded by observation noise $\B{F}$. As
with most fMRI ICA analysis procedures, we assume that the signal of
interest spans only a sub-space of the observation space. Components
$\B{C}$ spanning this subspace can be estimated using probabilistic
principal component analysis (PCA) \cite{beckmann2004}, which assumes
$\B{F}$ to be Gaussian-distributed, and lying in a subspace orthogonal to
$\B{C}$:
\begin{equation}
    \label{eqn:model3}
    \B{Y} = \B{W} \, \B{C} + \B{F},
\end{equation}
where $\B{Y}$ and $\B{F}$ are $(n_\text{time steps}, n_\text{voxels})$
matrices, the rows of which form pattern vectors. $\B{C}$ is the
$(n_\text{components}, n_\text{voxels})$ pattern matrix of the retained
principal components and $\B{W}$ is the matrix of their loadings in the
observed signal. In this paper, we do not discuss estimation of the
sub-space of interest, but focus on recovering sparse brain-activity
sources from $\B{C}$.

We model the patterns $\B{C}$ as generated by a set of sources $\B{A}$,
confounded by additive noise $\B{E}$, and observed as a linear mixture in
the sub-space spanned by $\B{C}$:
\stepcounter{equation}
\begin{align}
    \B{C} & = \B{M} \, \B{B}
    \qquad (2)
&
    \B{B} & = \B{A} + \B{E} 
    \label{eqn:model2}
\end{align}
$\B{M}$ is an orthogonal mixing matrix, $\B{A}$, $\B{B}$,
and $\B{E}$ are \linebreak$(n_\text{components}, n_\text{voxels})$ matrices.
Unlike $\B{F}$, $\B{E}$ is in the same sub-space as the brain sources. 
 In addition, we assume that the true sources correspond to the 
marginals $\B{A}_i$ that are sparse: most of the
coefficients of $\B{A}_i$ are zeros. As a result, the
histogram of $\B{A}_i$ is strongly super-Gaussian: it has heavy tails. If the amplitude of
the noise $\B{E}$ is small compared to the non-zero coefficients of $\B{A}$,
$\B{B}$ is also super-Gaussian and can be estimated from $\B{C}$ using
ICA. We use FastICA, a procedure that selects a basis of the signal
sub-space maximizing non-Gaussianity of the corresponding marginal
distributions \cite{Hyvarinen2000}.

If the components $\B{B}$ are observed mixed, the observed projections
$\B{C}_i$ reflect mostly the isotropic noise $\B{E}$ and not the sources
of interest $\B{A}$ that are sparse only in a particular basis. This is
why the estimation of the mixing model (2) is important for fMRI data
analysis, as the marginals on the estimated basis separate $\B{A}$ from
the background noise $\B{E}$.

{\bf Thresholding ICs to control for noise.}
We assume that the values of the non-zero voxels of sources $\B{A}$ are
larger than the standard deviation $\sigma$ of $\B{E}$. According to our model,
selecting voxels specific of the support of $\B{A}$ amounts to choosing a
threshold $\tau_\alpha$ to apply on the ICs $\hat{\B{B}}$.
ICA estimates particular directions of the feature space, thus a possible
null hypothesis $H_0$ for ICA is that all directions are equivalent. As a
result, the
null distribution for the marginals $\B{A}_i$ is given by
projections on random directions $\omega$ of the feature space. 
\begin{equation}
p(\B{A}_i > \tau_\alpha | H_0) =
\underset{\omega,\,||\omega||=1}{\text{mean}}p(|\omega^T \B{B}| > \tau_\alpha)
\end{equation}
We can sample this distribution directly from the data. In addition, $
\omega^T \B{B}$ is a linear combination of the random variables
$\B{B}_i$. As the sub-space has been whitened by the PCA, they all have a
variance of 1. For high dimensions, the central limit theorem thus states
that the distribution of $\omega^T \B{B}$ is Gaussian of variance 1. In
this case, the p-value is given by the inverse of the cumulative
distribution function of a
Gaussian process, and the threshold can be set as with a normal null.

\begin{figure}
    \includegraphics[width=0.32\linewidth]{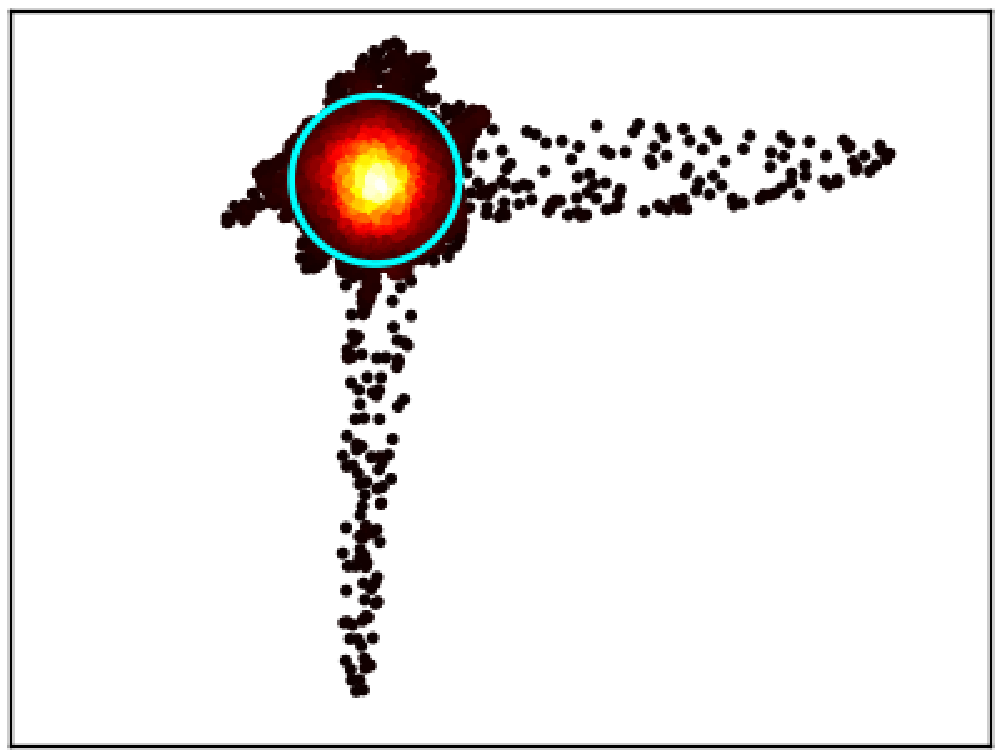}
    \llap{\raisebox{.3em}{\small\sffamily\bfseries a\;~}}%
    \hfill
    \includegraphics[width=0.32\linewidth]{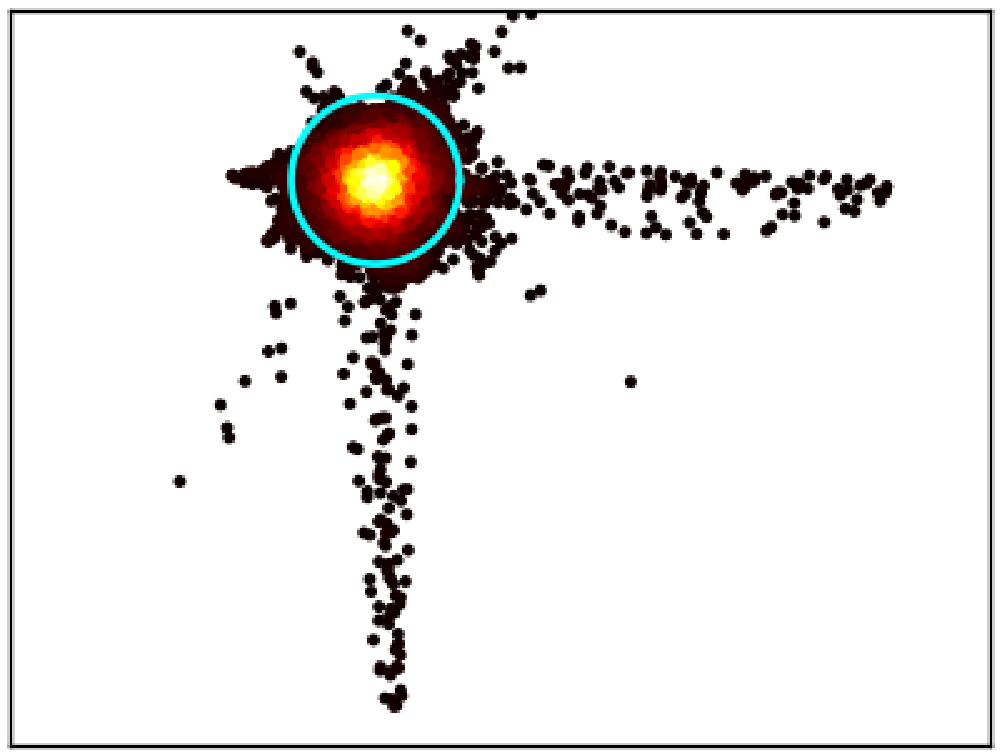}
    \llap{\raisebox{.3em}{\small\sffamily\bfseries b\;~}}%
    \hfill
    \includegraphics[width=0.32\linewidth]{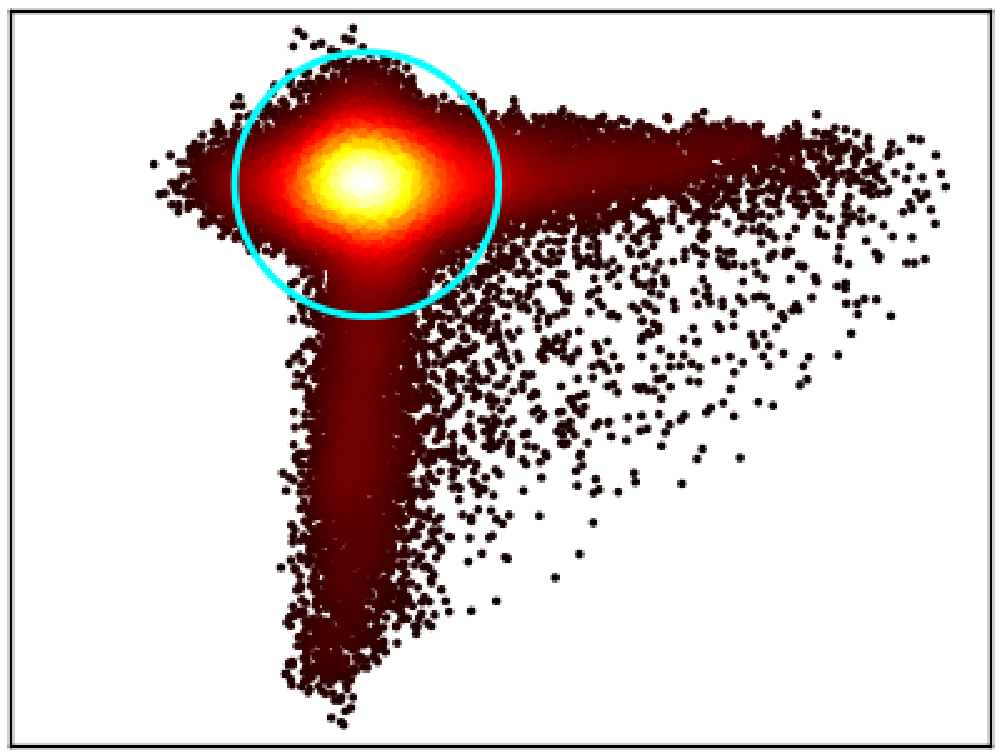}
    \llap{\raisebox{.3em}{\small\sffamily\bfseries c\;~}}%
    \vspace*{-2ex}

    \caption{
	\label{fig:feature_space}
	Scatter plot of samples projected in the subspace spanned by the
	two first ICs identified. The density is 
	represented by a colormap ranging from black (low density) to
	white (high density). The threshold as set by the model with 
	$p = 10^{-2}$ is represented as a light blue circle.
	{\sffamily\bfseries (a)} Simulated data with 9 features total, 
	with $\B{E}$ generated from a Gaussian process with
	$\sigma = 0.15$. 
	{\sffamily\bfseries (b)} Same simulations with super-Gaussian noise 
	(kurtosis of $4$). 
	{\sffamily\bfseries (c)} fMRI data.
}
\end{figure}

A representation of the signal in feature space is given on 
Fig.~\ref{fig:feature_space} for various distributions: synthetic data
generated from the model exposed above (Fig.~\ref{fig:feature_space}a),
synthetic data with additional super-Gaussian noise, 
(Fig.~\ref{fig:feature_space}b), and fMRI data 
(Fig.~\ref{fig:feature_space}c).
All share a central mode corresponding to $\B{E}$ in our description,
that can be approximated as a multivariate Gaussian process. In addition,
for each mixing direction, activated voxels can be found when moving away
from the center.

Our model is different from most noisy ICA models, as they assume that
contribution of the noise to the signal sub-space is small. They account
for the noise in the ICA estimation by correcting the bias it introduces
to the whitening and the measures of statistical independence \cite{cichocki1998}. In
our model, noise accounts for a large fraction of the variance in the
signal sub-space.


\section{Simulation study}

\label{sec:simulation_study}

We generate synthetic samples $\tilde{\B{Y}}$ from our model with a known
ground truth and noise model. We consider 9 features $\tilde{\B{A}}$,
that is 2D maps (80, 80) pixel large and made of one or two rectangles of
uniformly-active pixels on a null background. We add random noise
$\tilde{\B{E}}$ generated by a multivariate normal distribution of
isotropic variance 1. We control the amplitude of the noise with a
parameter $\lambda$: \mbox{$\tilde{\B{B}} = \tilde{\B{A}} + \lambda
\tilde{\B{E}}$}. We draw a random rotation matrix $\tilde{\B{M}}$ to mix
the patterns $\tilde{\B{B}}$, and apply a Gaussian spatial smoothing of
FWHM 2 pixels to simulation the point spread function of the scanner. Due
to the smoothing, the noise term is observed as a random Gaussian field
with a reduced variance compared to the initial random process. We set
$\lambda$ to control the variance of this field. 

In addition, as it is likely that, in real fMRI settings, not all
background noise can be described by Gaussian processes, we
generate synthetic data with non-Gaussian noise. For this, in addition 
to the previous Gaussian random field, 
$\tilde{\B{E}}_\text{g}$, we generate a super-Gaussian contribution
$\tilde{\B{E}}_\text{ng}$ by applying a non-linear rescaling to a
smoothed Gaussian random field. We use the cubic non-linearity that
generates {\sl spiky} noise with a long-tailed distribution. The
additional noise term is thus sparse, and not invariant by rotation of
the feature space. We add it to the signal of interest in the observation
basis. We set the contributions of both noise terms to control the
variance and kurtosis of the resulting random process: $\tilde{\B{C}} =
\tilde{\B{M}}\,\tilde{\B{A}} + \lambda\, (\cos\theta \,
\tilde{\B{M}}\,\tilde{\B{E}}_\text{g} + \sin\theta \,
\tilde{\B{E}}_\text{ng})$. This structured noise term violates the noise
model of the ICA algorithm and poses thus a challenge to the feature
extraction by offsetting the estimation of the mixing matrix, and thus
the projection.

\begin{figure}
    \includegraphics[width=\linewidth]{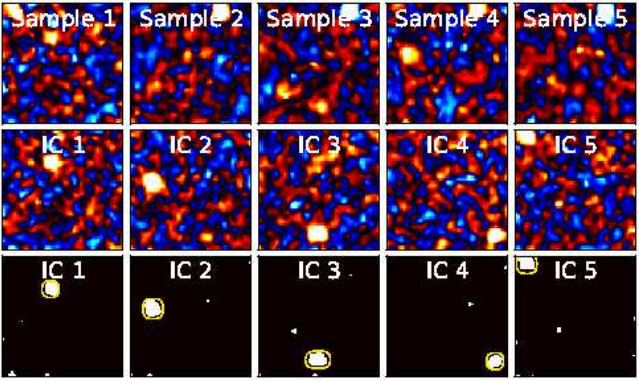}
    \vspace*{-5ex}%
    \caption{
    \label{fig:synthetic_data} Simulated data, showing 5 samples out of
9, for $\B{E}$ generated from a super-Gaussian process with
$\sigma = 0.15$ and a kurtosis of $4$. 
The threshold is set by the model with $p = 10^{-2}$. {\bf Top row}: 
    observed samples $\B{Y}$. 
    {\bf Middle row}: ICs $\B{B}$.
    {\bf Bottom row}: estimated sources $\B{A}$, the ground truth is 
    outlined in light yellow.
    }
\end{figure}

Spatial maps generated by the simulations are presented on 
Fig.~\ref{fig:synthetic_data}. The samples projected in feature space on the 2
first ICs are presented on Fig.~\ref{fig:feature_space}. We apply ICA
estimation and thresholding as described above.
To quantify the
specificity and the sensitivity in feature detection, we plot
receiver-operator characteristics on Fig.~\ref{fig:roc_plot} for Gaussian
and super-Gaussian (kurtosis $= 4$) noise. Increasing noise amplitude
$\sigma$ degrades estimation performance, as the central mode becomes
indistinguishable from the outliers we are interested in. Performances
are slightly degraded by the addition of the super-Gaussian noise. It
induces errors in the choice of the projection basis, as can be seen on
Fig.~\ref{fig:feature_space}b: in the projected space, sources are not
completely unmixed. In addition, on Tab. \ref{tab:error_rates}, we
compare false positive rates to the specified p-value. We find that for
Gaussian noise amplitudes up to $\sigma = 0.20$ or super-Gaussian noise
amplitude of $\sigma = 0.15$, the p-values give an exact control on
type 1 errors. With more noise, the tail of the central mode cannot
account for all false detections for small p-values. We stipulate that
the additional errors come from projection error due to incomplete
source unmixing by the ICA procedure.

\begin{table}
{\small
\begin{center}
\begin{tabular}{r|ccc}
Specified p-value & $5{\cdot}10^{-2}$ & $1.0{\cdot}10^{-2}$ & $5.0{\cdot}10^{-3}$
\\
\hline
Gaussian, $\sigma=.15$ 	      & $4.0{\cdot}10^{-2}$ &  $7.1{\cdot}10^{-3}$ &  $4.0{\cdot}10^{-3}$
\\
super-Gaussian, $\sigma=.15$  & $4.2{\cdot}10^{-2}$ &  $1.0{\cdot}10^{-2}$ &     $6.2{\cdot}10^{-3}$
\\
Gaussian, $\sigma=.20$ 	      & $4.9{\cdot}10^{-2}$ &  $9.4{\cdot}10^{-3}$ &  $5.2{\cdot}10^{-3}$
\\
super-Gaussian, $\sigma=.20$  & $5.2{\cdot}10^{-2}$ &  $1.3{\cdot}10^{-2}$ &     $7.9{\cdot}10^{-3}$
\\
Gaussian, $\sigma=.30$ 	      & $6.0{\cdot}10^{-2}$ &  $1.3{\cdot}10^{-2}$ &  $7.4{\cdot}10^{-3}$
\\
super-Gaussian, $\sigma=.30$  & $5.9{\cdot}10^{-2}$ &  $1.5{\cdot}10^{-2}$ &     $1.0{\cdot}10^{-2}$
\\
\hline
fMRI data                     & $3.6{\cdot}10^{-2}$ &  $1.7{\cdot}10^{-2}$ &     $1.3{\cdot}10^{-2}$
\end{tabular}
\end{center}
}%
\vspace*{-1.5em}%
    \caption{
    \label{tab:error_rates} False positive rates as a function of 
    model-based p-value, for simulated and fMRI data. 
    }
%
%
\end{table}

\section{fMRI study} 

\label{sec:fmri_study}

\begin{figure*}
\def\pagefraction{.31}
\begin{tabular}{p{\pagefraction\linewidth}|%
	        p{\pagefraction\linewidth}|%
	        p{\pagefraction\linewidth}}
    \hspace*{-.025\linewidth}%
    \includegraphics[width=1.05\linewidth]{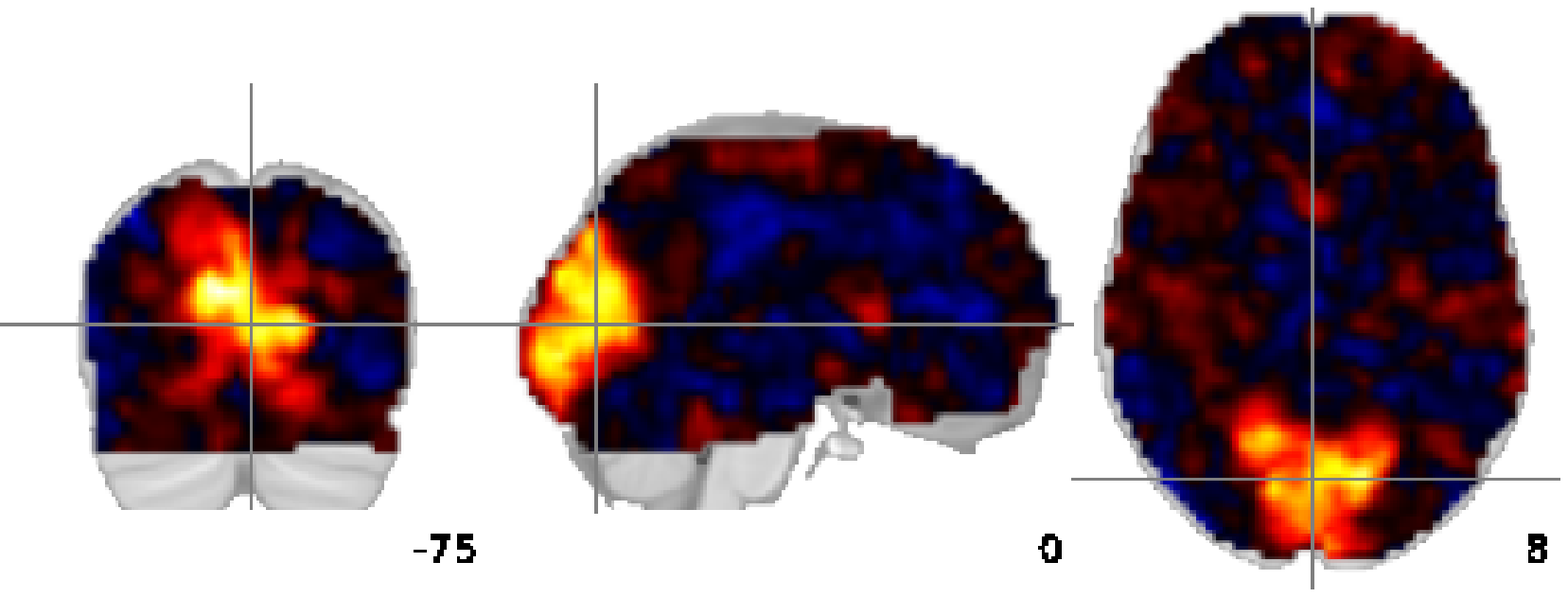}
    &
    \hspace*{-.025\linewidth}%
    \includegraphics[width=1.05\linewidth]{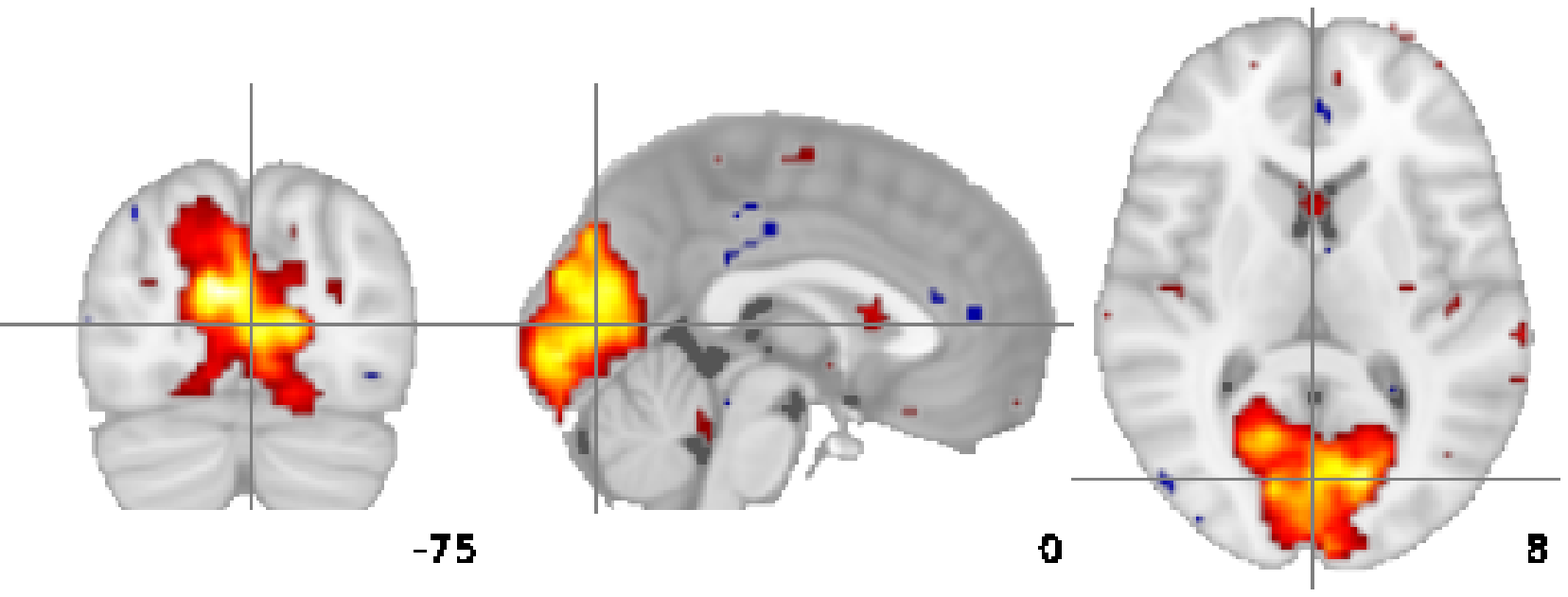}
    &
    \hspace*{-.025\linewidth}%
    \includegraphics[width=1.05\linewidth]{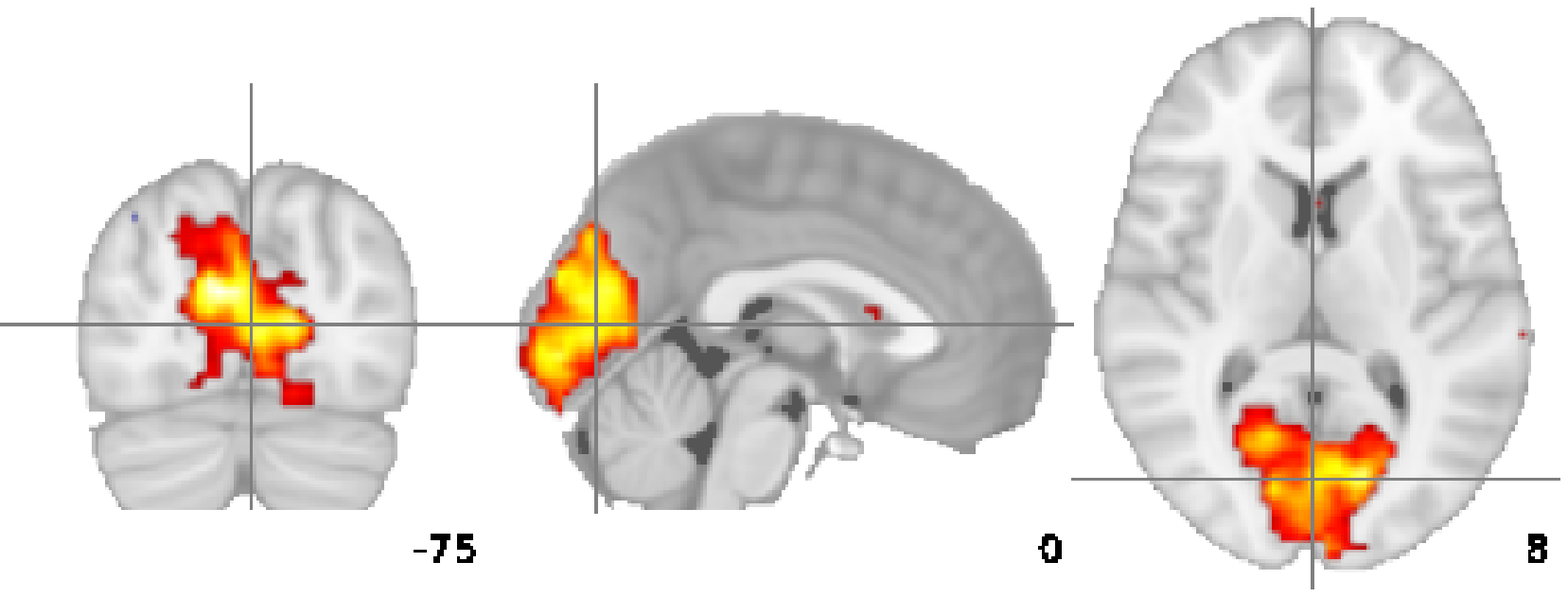}
\\[-6.7em]
    \sf\small Pseudo ground truth
    &
    \sf\small 
    {\footnotesize MELODIC mixture model}
    &
    \sf\small 
    \sf{\footnotesize Multivariate thresholding procedure}
\\[5.6em]
    \hspace*{-.025\linewidth}%
    \includegraphics[width=1.05\linewidth]{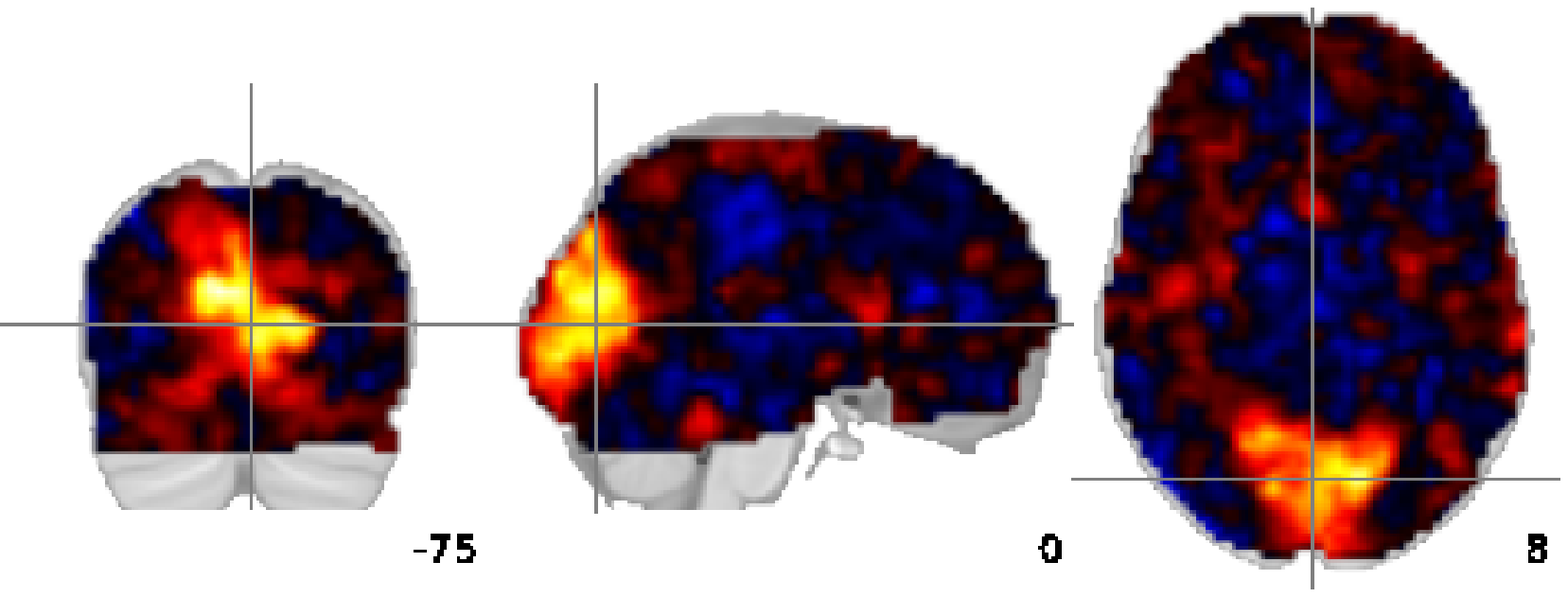}
    &
    \hspace*{-.025\linewidth}%
    \includegraphics[width=1.05\linewidth]{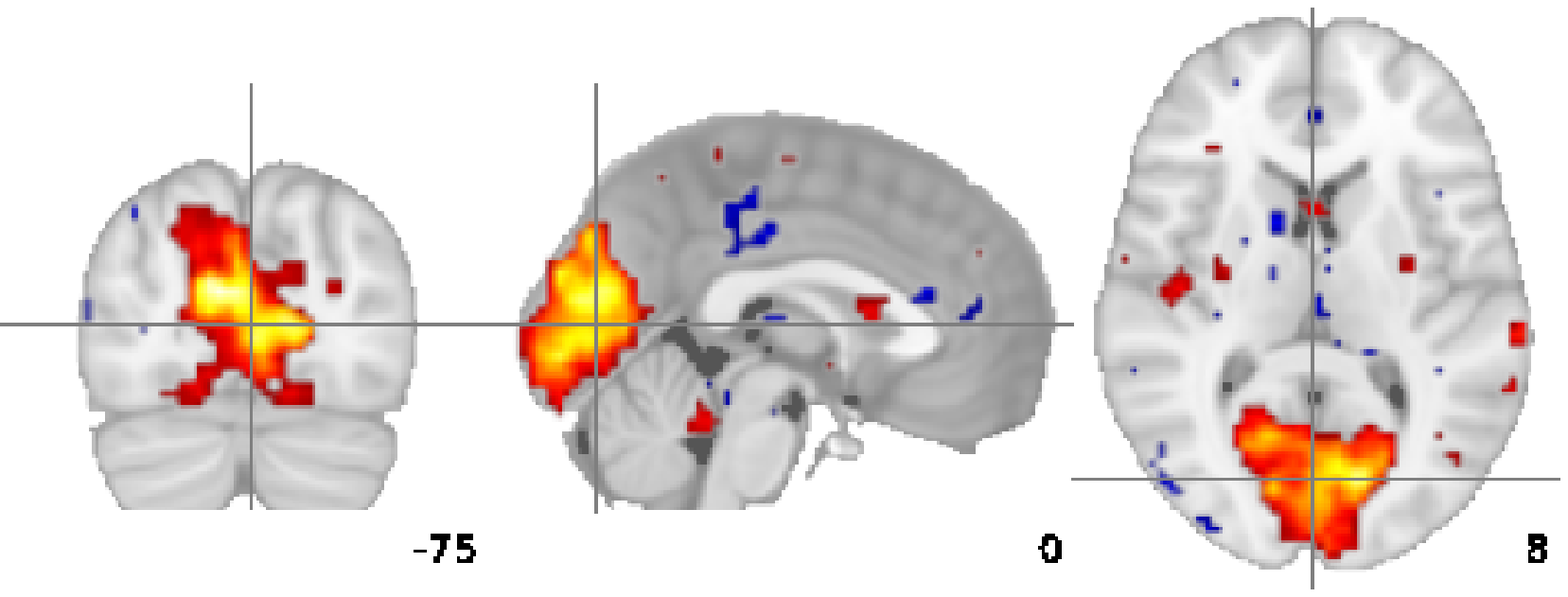}
    &
    \hspace*{-.025\linewidth}%
    \includegraphics[width=1.05\linewidth]{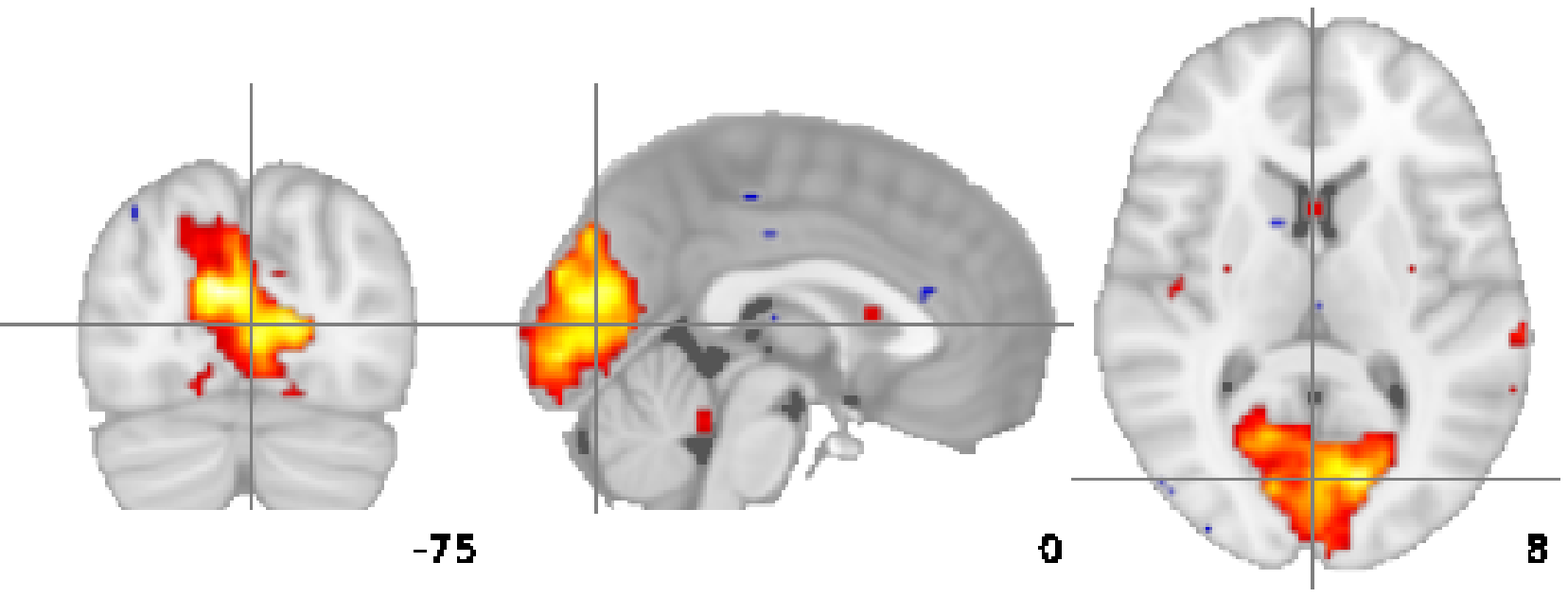}
\\[-6.7em]
    \sf\small Downsampled data
    &
    \sf\small 
    &
    \sf\small 
\\[5.6em]
\end{tabular}
\rule{\linewidth}{1pt}
\begin{tabular}{p{\pagefraction\linewidth}|%
	        p{\pagefraction\linewidth}|%
	        p{\pagefraction\linewidth}}
    \hspace*{-.025\linewidth}%
    \includegraphics[width=1.05\linewidth]{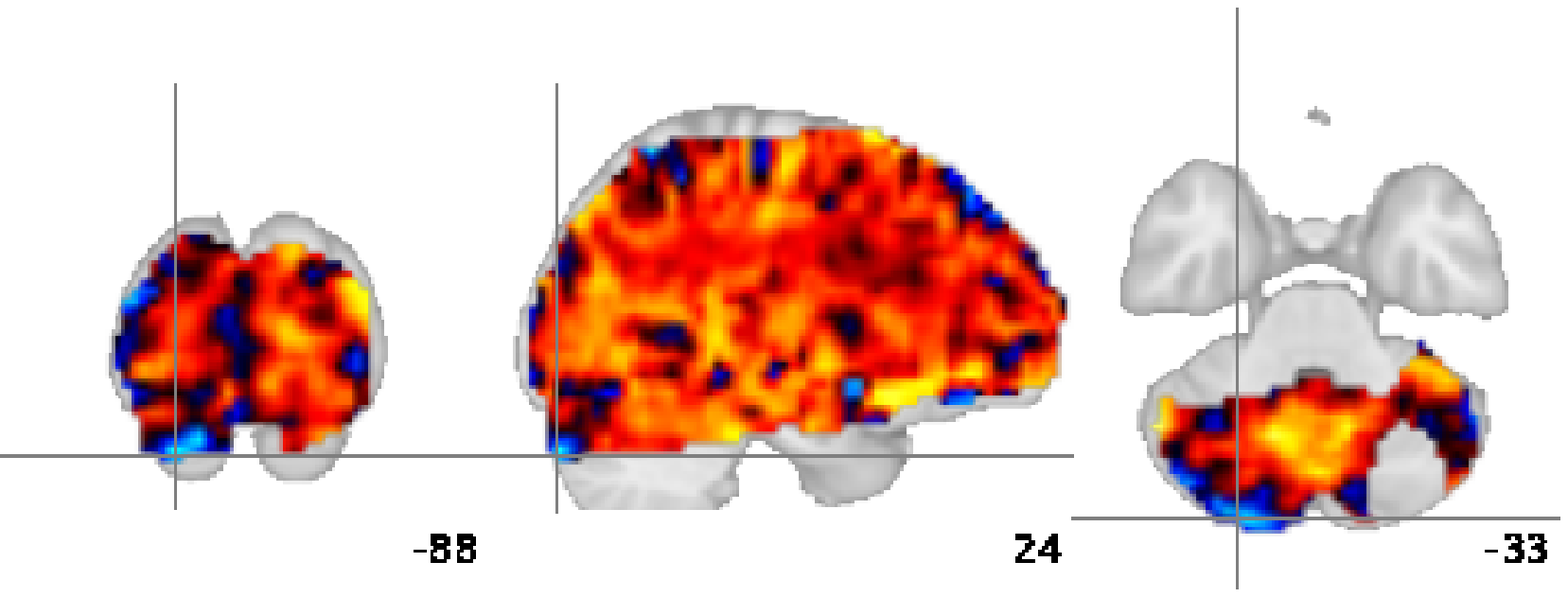}
    &
    \hspace*{-.025\linewidth}%
    \includegraphics[width=1.05\linewidth]{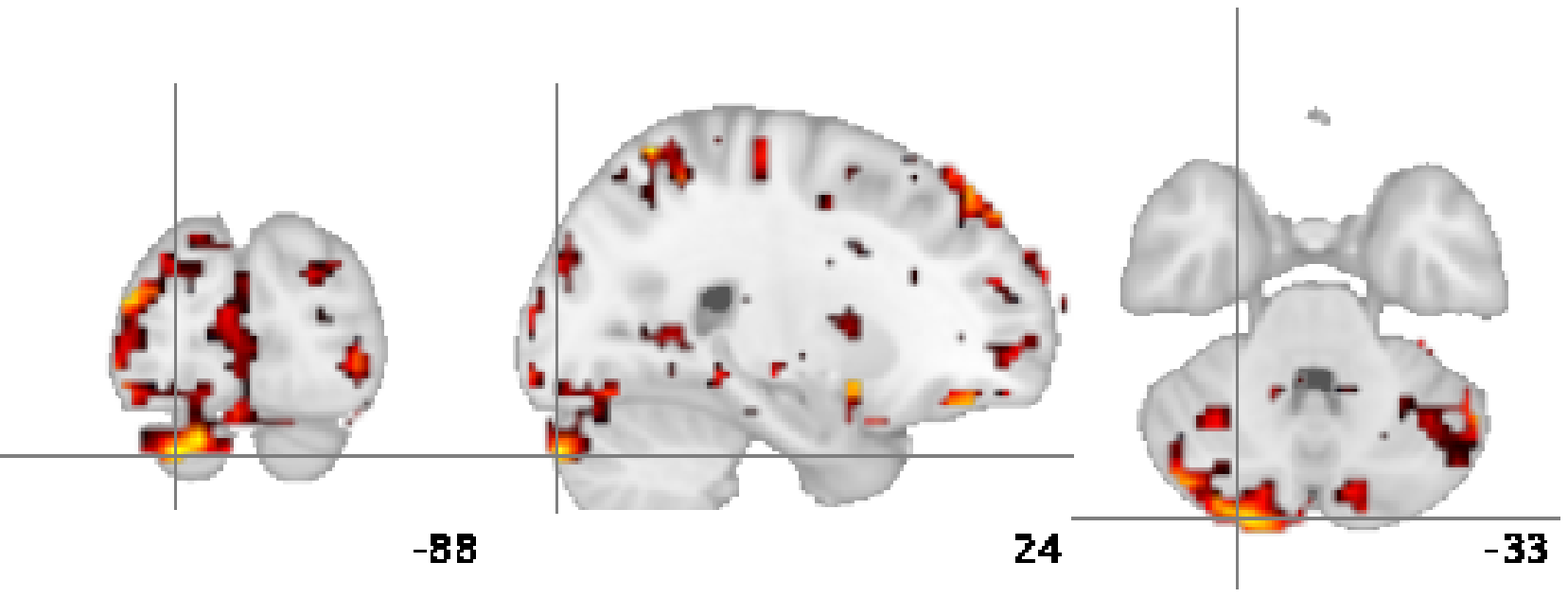}
    &
    \hspace*{-.025\linewidth}%
    \includegraphics[width=1.05\linewidth]{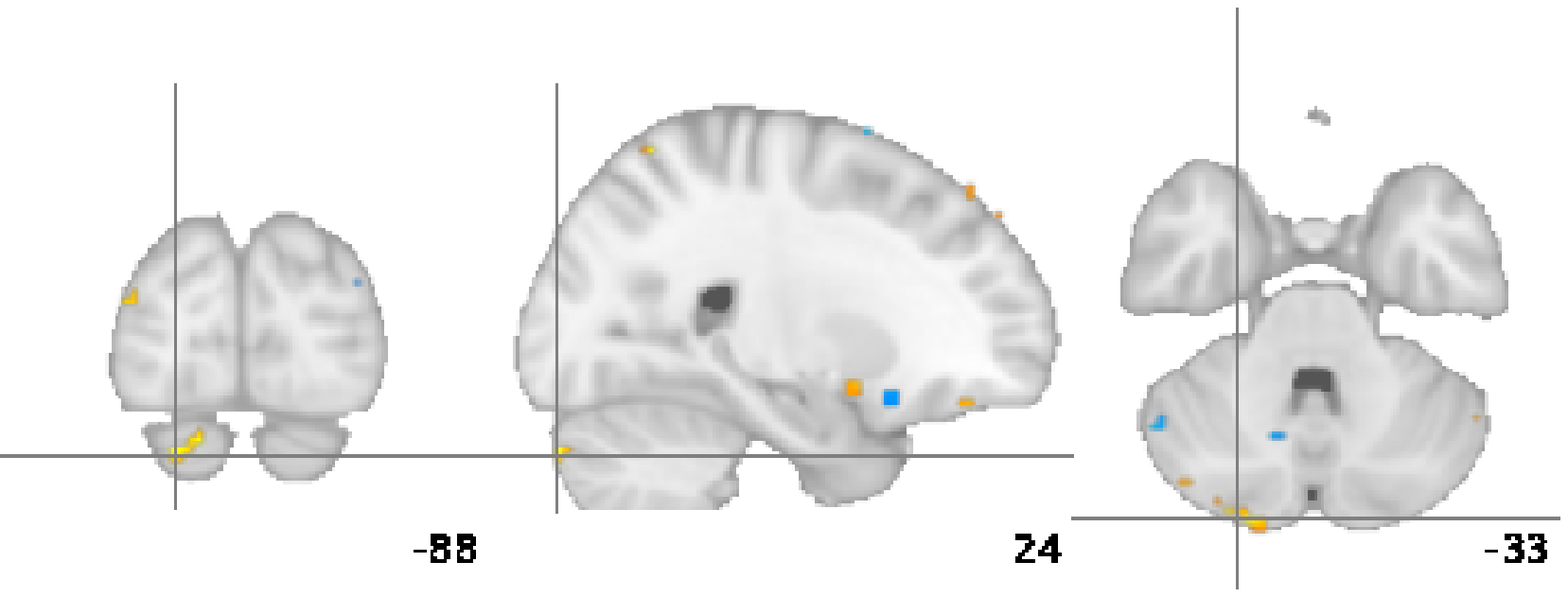}
\\[-6.7em]
    \sf\small Pseudo ground truth
    &
    \sf\small 
    {\footnotesize MELODIC mixture model}
    &
    \sf\small 
    \sf{\footnotesize Multivariate thresholding procedure}
\\[5.6em]
    \hspace*{-.025\linewidth}%
    \includegraphics[width=1.05\linewidth]{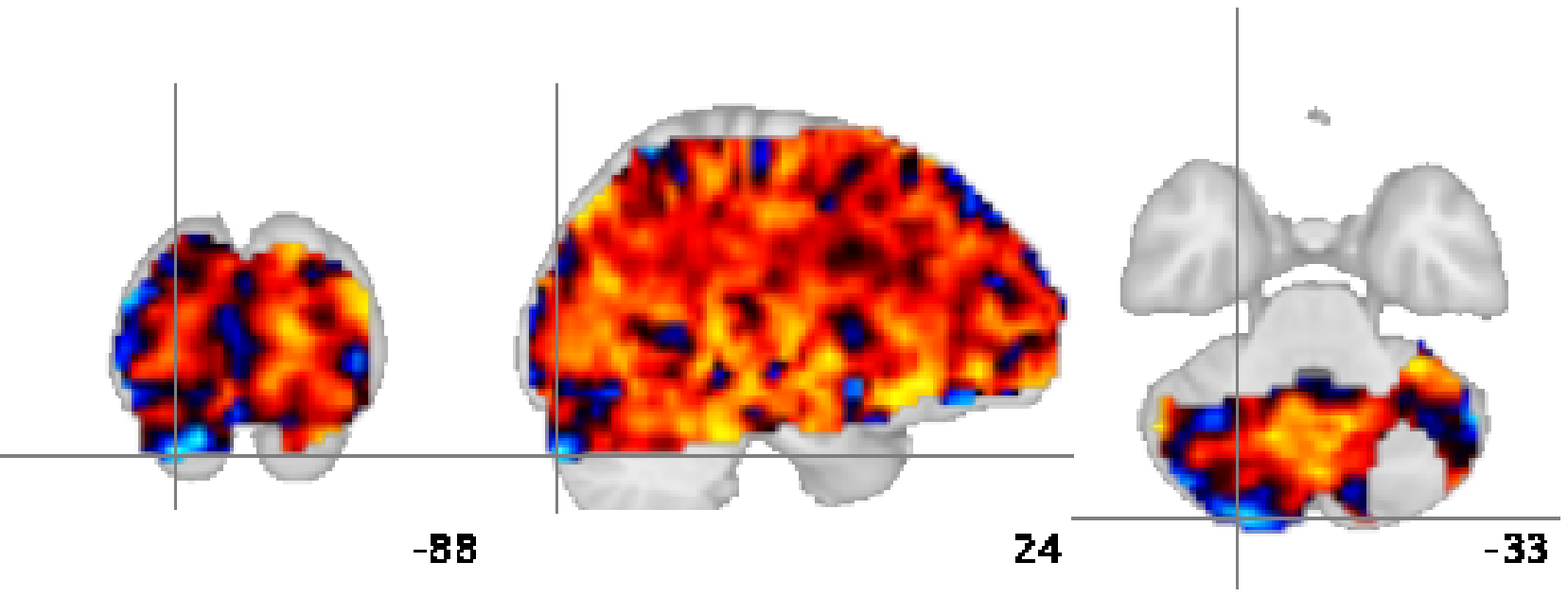}
    &
    \hspace*{-.025\linewidth}%
    \includegraphics[width=1.05\linewidth]{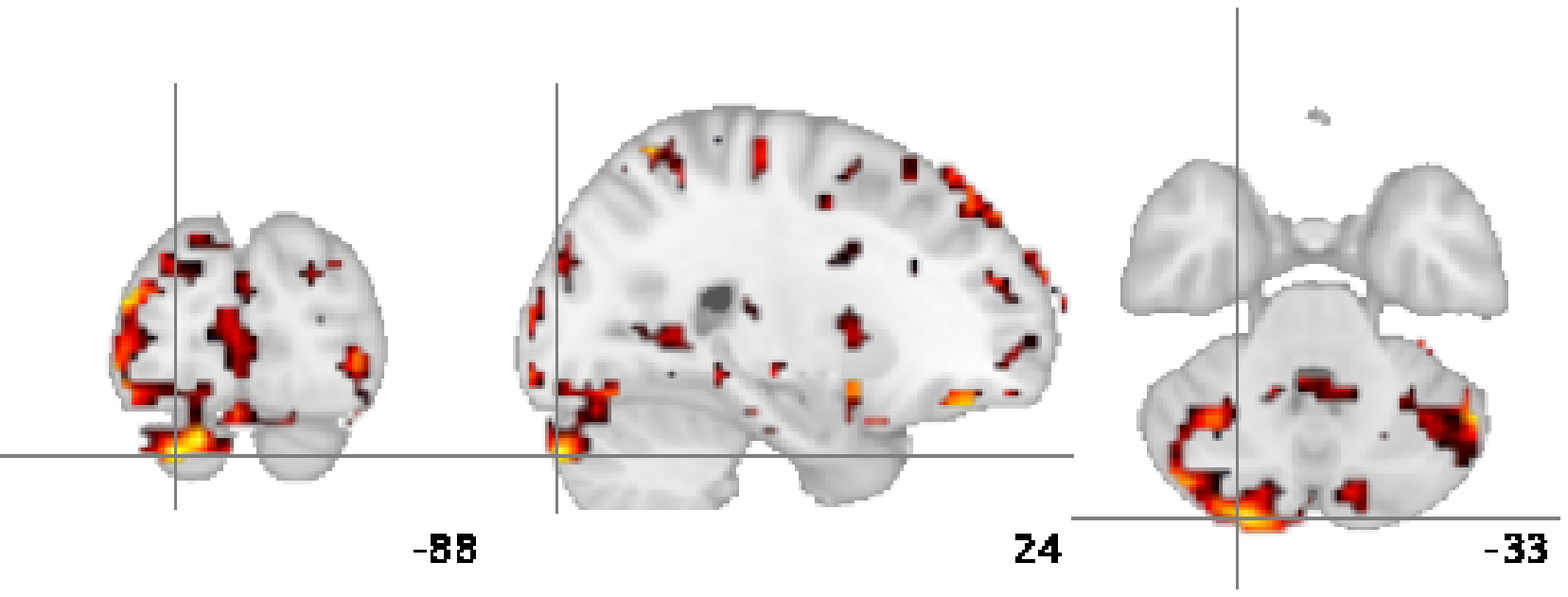}
    &
    \hspace*{-.025\linewidth}%
    \includegraphics[width=1.05\linewidth]{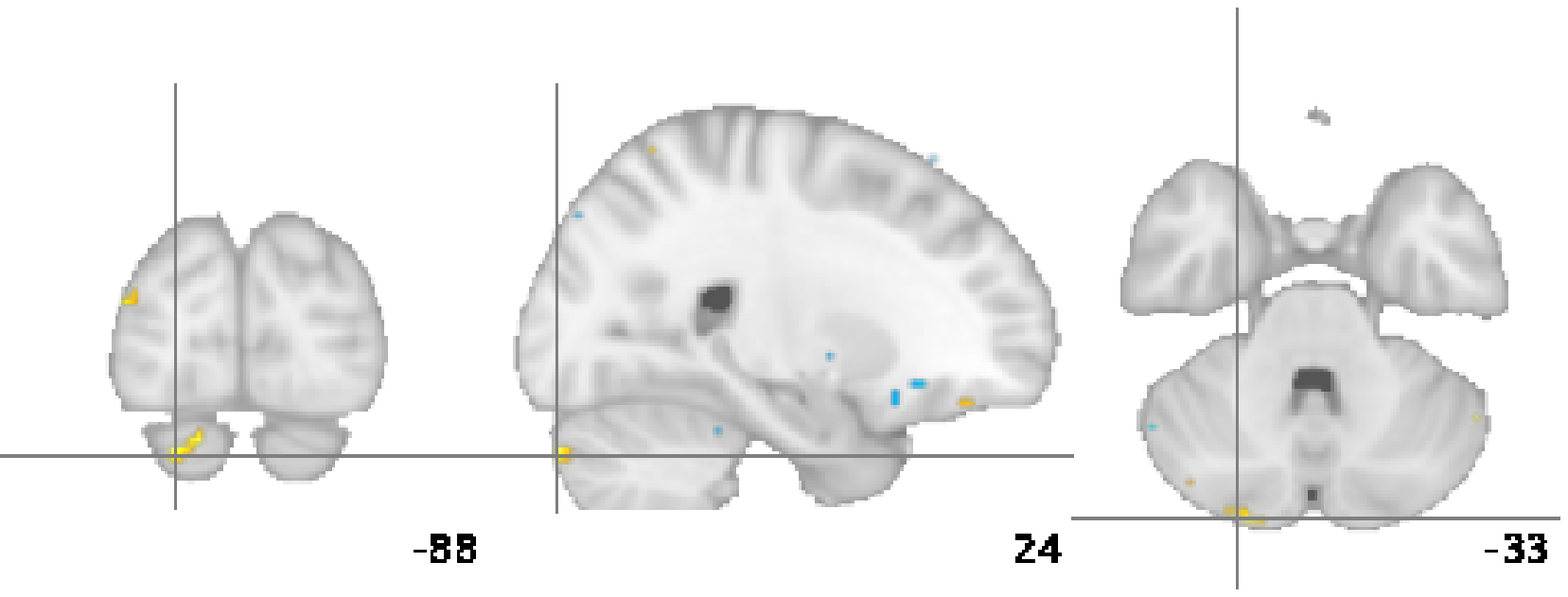}
\\[-6.7em]
    \sf\small Downsampled data
    &
    \sf\small 
    &
    \sf\small 
\\[5.6em]
\end{tabular}
\caption{
    \label{fig:fmri_data} ICs estimated from fMRI data
and thresholded using MELODIC's mixture model, and our multivariate
thresholding procedure. {\bf Top rows:} IC detecting the primary visual areas.
{\bf Bottom rows:} IC representative of a vascular artifact.
}
\end{figure*}

We apply our method to fMRI data for 12 subjects at rest from a previous
study \cite{sadaghiani2009}. 820 volumes were acquired with a repetition
time (TR) of $1.5\,$s. We run the procedure (ICA analysis and
thresholding) for single-subject data on the first 40 principal
components. For fMRI data, the ground truth is not known, so we generate
degraded datasets from the original dataset, and consider the latter as a
pseudo ground truth to quantify error rates. This procedure quantifies
consistency of the estimator in the presence of noise. To generate
degraded datasets while retaining observations of the same brain
activity, we use one volume out of 3. The effective TR of the
down-sampled datasets is $4.5\,$s. This sampling rate is enough to retain
most of the hemodynamic response, convolved by the 6-second-long response
function. In addition, the 3 resulting interleaved time series sample
different high-frequency noise that confounds the signal of interest.
Thresholded ICs estimated on the various resampled datasets for one
subject are matched with the corresponding pseudo ground truth. 
Fig.~\ref{fig:fmri_data} presents pseudo ground truth and downsampled data. On
non-thresholded ICs, we can see that the level of background noise is
indeed higher in ICs learned on downsampled data. We run the MELODIC
mixing model on the ICs to compare sensitivity (false negatives) and
specificity (false positives).

\begin{figure}
    \begin{center}
    \includegraphics[width=.8\linewidth]{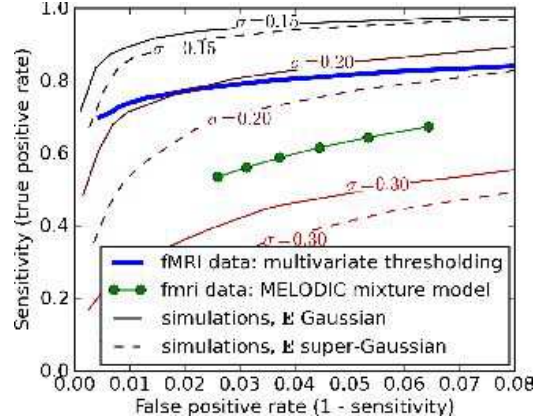}
    \vspace*{-4ex}%
    \end{center}
    \caption{
    \label{fig:roc_plot} ROC plot: sensitivity as a function of
    false positive rate for synthetic data using Gaussian and 
    super-Gaussian (kurtosis = 4) noise of varying $\sigma$, 
    as well as for fMRI data.
    }
\end{figure}

As seen on the ROC plot (Fig.~\ref{fig:roc_plot}), average performance on
fMRI data for the 12 subjects is on par with simulated data. Good control
of false positives can be achieved, but the true positive rate remains
limited. This can be explained by errors in our pseudo-ground truth. In
addition, the false positive rate is controlled by the specified p-value
only to $10^{-2}$, although to account for errors in the pseudo-ground
truth, the observed false positive rate should be corrected by a factor
$0.5$. With MELODIC's mixture model, we specify different inter-class
mixing probability ratios to vary specificity; we do not report on very 
large or very small ratios as they induce non-monotonous thresholding and
poor overall performance.
Our multivariate thresholding proceeding can achieve better
specificity/sensitivity trade off MELODIC's mixture model. 

ICs estimated on fMRI data most often display a few salient features
related to anatomical regions and may be interpreted as brain networks.
On such IC, both our thresholding procedure and MELODIC's mixture model
extract similar regions, although our procedure yields fewer
small clusters outside of the main segmented areas (see 
Fig.~\ref{fig:fmri_data}, top). In contrast, some ICs, representative of
non-cognitive processes such as blood flow or movement, are very
fragmented and diffuse with no region strongly standing out. On these
ICs, a mixture model fits the null distribution to the center of the
histogram, and thus selects large regions, whereas our thresholding
procedure selects very few voxels, as it does not consider the component
by itself, but as part of the complete multivariate signal (see 
Fig.~\ref{fig:fmri_data}, bottom).

\section{Conclusion}
\label{sec:conclusion}

This contribution presents a procedure for thresholding ICA patterns of
fMRI time series to recover sparse sources using a
multivariate model of spatially-sparse brain activity that does not rely
on correlating with external stimuli. From a practical point of view, the
main improvement over existing ICA-based methods for fMRI is that
non-neuronal patterns are rejected as they do not correspond to very salient 
features. We have validated on simulated data and
resting-state fMRI data that the procedure can yield exact control of the
false positive rates for $p>10^{-2}$ and achieves better
sensitivity/specificity trade-offs than the current state-of-art fMRI ICA
support-selection procedures. Control of false
detections and consistency of estimation on noisy data is important for
clinical and medical research applications of resting-state fMRI. Our
procedure can be understood as outlier detection with projection pursuit,
as proposed by Gnanadesikan and Kettenring \cite{gnanadesikan1972}, using
ICA.

\bibliographystyle{IEEEbib_short}
\bibliography{restingstate}

\end{document}